\documentclass[a4paper,twocolumn,superscriptaddress,pre,10pt]{revtex4-2}
\usepackage{float} 
\usepackage{bm,amsmath}
\usepackage{graphicx}
\usepackage{hyperref}
\usepackage{xcolor}
\usepackage{multirow}
\usepackage{array}
\usepackage{makecell}
\usepackage{comment}

\newcolumntype{C}[1]{>{\centering\let\newline\\\arraybackslash\hspace{0pt}}m{#1}}

\hypersetup{
    colorlinks,
    linkcolor={red!90!black},
    citecolor={blue!90!red},
    urlcolor={blue!100!black}
}

\newcommand{\f}{\frac}
\renewcommand{\d}{\mathrm{d}}

\begin{document}

\title{Resetting dynamics in a system with quenched disorder}
\author{Riya Verma}
\affiliation{Tata Institute of Fundamental Research, Gopanpally Village, Hyderabad - 500046, India}
\author{Binayak Banerjee}
\affiliation{Tata Institute of Fundamental Research, Gopanpally Village, Hyderabad - 500046, India}
\author{Shamik Gupta}
\email{shamik.gupta@theory.tifr.res.in}
\affiliation{Tata Institute of Fundamental Research, Mumbai, India}
\author{Saroj Kumar Nandi}
\email{saroj@tifrh.res.in}
\affiliation{Tata Institute of Fundamental Research, Gopanpally Village, Hyderabad - 500046, India}

\begin{abstract}
Although resetting has widespread applicability, applying it to the dynamics in the presence of spatial quenched disorder, which is essential in many physical problems, is challenging. In this study, we consider a well-known one-dimensional model of particle hopping on a lattice with quenched disorder in the form of site-dependent hopping probabilities, drawn from a power-law distribution, and apply the resetting formalism. As a physical example, we recast the growth dynamics of microtubules with sudden catastrophic disassembly events as a resetting dynamics. We consider two distinct regimes for growth dynamics: a strongly biased case and a less biased case. Motivated by experimental results, we take a Gamma distribution for the resetting time. Our results show that occasional disassembly events are crucial for the experimentally observed distribution of reset (or catastrophe) lengths. We also analyze steady-state distributions under different resetting protocols—resetting to the initial position versus a random site. We also investigate the distribution of first-passage times to a fixed distance following reset. Finally, by considering other resetting probability distributions, we identify a regime where the mean displacement grows as slowly as $\log^2 t$. We also elucidate the role of disorder in the system properties under the resetting dynamics. Our study paves the way to treat the dynamics of complex physical systems using resetting.

\end{abstract} 
\maketitle

\section{Introduction}
Resetting a particle to its initial, or some other randomly chosen, conditions can be applied to any stochastic process \cite{evans2011,evans2020}. It provides a generic framework for studying a wide variety of problems. Examples include different variants of Brownian motion \cite{gupta2019stochastic, santra2022effect,majumdar2018}, diffusive particles \cite{evans2020}, run and tumble particles \cite{santra2020run}, colloidal systems \cite{tal2020experimental}, optimization of search processes by minimizing mean target-finding times \cite{evans2020, gupta2022stochastic}, homing processes \cite{paramanick2024}, etc. In addition, resetting is a means to drive the system out of equilibrium, leading to various steady states. Over the last decade or so, the process has been applied to diverse scenarios, such as non-Poissonian resetting, interacting particles, and memory effects \cite{evans2020}, as well as diffusion in potential landscapes \cite{pal2016diffusion}. Moreover, there are examples of natural processes that evolve slowly and get interrupted by sudden changes or resets: At low shear rates and low temperatures, amorphous solids exhibit intermittent plastic flow — seen as jerky drops in stress \cite{antonaglia2014bulk}. Earthquake cycles show irregular stress accumulation and sudden release \cite{brace1966,vere1988variance}. In foraging, animals optimize target finding via alternate random search and fast ballistic motion \cite{benichou2005optimal}. In population dynamics, immigration–birth–death models are occasionally hit by total catastrophes that lead to abrupt population decline \cite{kyriakidis1994stationary}.

The sudden changes in specific observables in these problems suggest that applying the resetting framework can yield crucial insights. However, in practice, such an application is not straightforward. Analytical approaches to understanding the effects of resetting typically assume knowledge of the process in its absence. This primary assumption is often nontrivial for most physical processes, even in their simplest realizations, due to the presence of disorder. The properties of disordered systems have long been a central topic in statistical physics, as they control transport, relaxation, and localization across diverse physical systems \cite{gaspard2015cycles, gopalakrishnan2020dynamics, anderson1958absence,vojta2019disorder, bouchaud1990}, and interesting physical characteristics emerge with varying degrees of disorder in systems, such as polymer electrolytes, molten salts, disordered metal hydrides, and glassy systems  \cite{price2003dynamical}. However, no generic framework for properties in the presence of disorder exists to date. Exact analytical results exist only for a few exceptional cases, such as the quenched disorder in one spatial dimension. For example, a static disorder can lead to non-Gaussian propagators and a breakdown of self-averaging in simple random hopping models \cite{Bernasconi1978}. Disorder induces ultra-slow diffusion, with particle displacement scaling as  $ \langle x(t) \rangle \sim (\ln t)^2 $ and mean-square displacement $\langle x^2(t) \rangle \sim (\ln t)^4 $ in the Sinai model \cite{sinai1982lecture, marinari1983random}. One can also calculate the velocity and diffusion constant as explicit functions of disorder parameters \cite{derrida1982}. The disorder, in the form of a broad trapping-time distribution in a trap model \cite{bouchaud1990,Bouchaud1992}, can lead to weak ergodicity breaking and aging in disordered systems.

In addition, there are examples from biological systems where resetting-like dynamics is quintessential. In these systems, disorder often plays a crucial role. For example, in gene expression, ribosome movement along mRNA during translation can be modeled as transport on a one-dimensional lattice with quenched disorder, where site-dependent hopping rates are determined by tRNA availability \cite{shaw2004mean}. Interestingly, during transcription, RNA polymerases also undergo backtracking—a form of 1D diffusion with resetting-like dynamics that enable resumption of RNA synthesis \cite{roldan2016stochastic}. The growth trajectory of microtubules also shows sudden disassembly of the monomers or catastrophe \cite{gardner2011depolymerizing}; this closely resembles the dynamics under resetting. Microtubule growth generally depends on the local environment \cite{margolin2012mechanisms, schek2007microtubule}, which leads to heterogeneity in growth. Actin filaments also show sharp catastrophe-like events known as severing \cite{pavlov2007actin, roland2008stochastic}. Severing proteins bind to filaments, inducing breakage and fragmenting the filament, leading to resetting-like dynamics.

In this work, we apply the resetting framework to a one-dimensional lattice model with quenched disorder in the hopping probabilities. Instead of focusing on a specific problem, our aim here is to demonstrate that the resetting framework can provide critical insights into a diverse class of physical problems with disorder. However, for concreteness, we use a specific problem as motivation and explore three scenarios to demonstrate the formalism's applicability. We study a one-dimensional hopping model with quenched disorder in the hopping probabilities, drawn from a well-known power-law distribution. The first problem of interest is the growth dynamics of microtubules, with sudden catastrophic disassembly, which we treat as a reset within our formalism. Motivated by experiments \cite{gardner2011depolymerizing},  we draw the resetting times from a Gamma distribution. We then use exponentially and power-law distributed resetting times and show that resetting can either stop growth or lead to anomalously slow growth dynamics. Similar anomalously slow growth dynamics can appear in physical systems, for example, growing bacterial colonies in a nutrient-limited environment \cite{bren2013} or disordered glassy systems under pinning \cite{saroj2014}.

We have studied two different resetting schemes: resetting to its initial position and to a randomly chosen previously visited site. For both resetting schemes, we examine the distribution of reset lengths under a Gamma-distributed resetting time. We also analyze the first-passage time (FPT) distribution for the particle to reach and return to a fixed distance $d$. This distribution exhibits a power-law decay at short times and an exponential cutoff at long times. We find that resetting can induce a steady-state distribution of the particle's displacement. The rest of the paper is organized as follows: In Sec. \ref{model}, we introduce the model, including the nature of disorder and resetting times. We present the results in Sec. \ref{result}. 
We first show the dynamics under resetting, motivated by the problem of microtubule growth, in Sec. \ref{microtubule}, and then with various other resetting protocols in Sec. \ref{otherreset}. We conclude the paper in Sec. \ref{disc} with a discussion of our results.

\section{The model}
\label{model}
\begin{figure}
	\includegraphics[width=8.6cm]{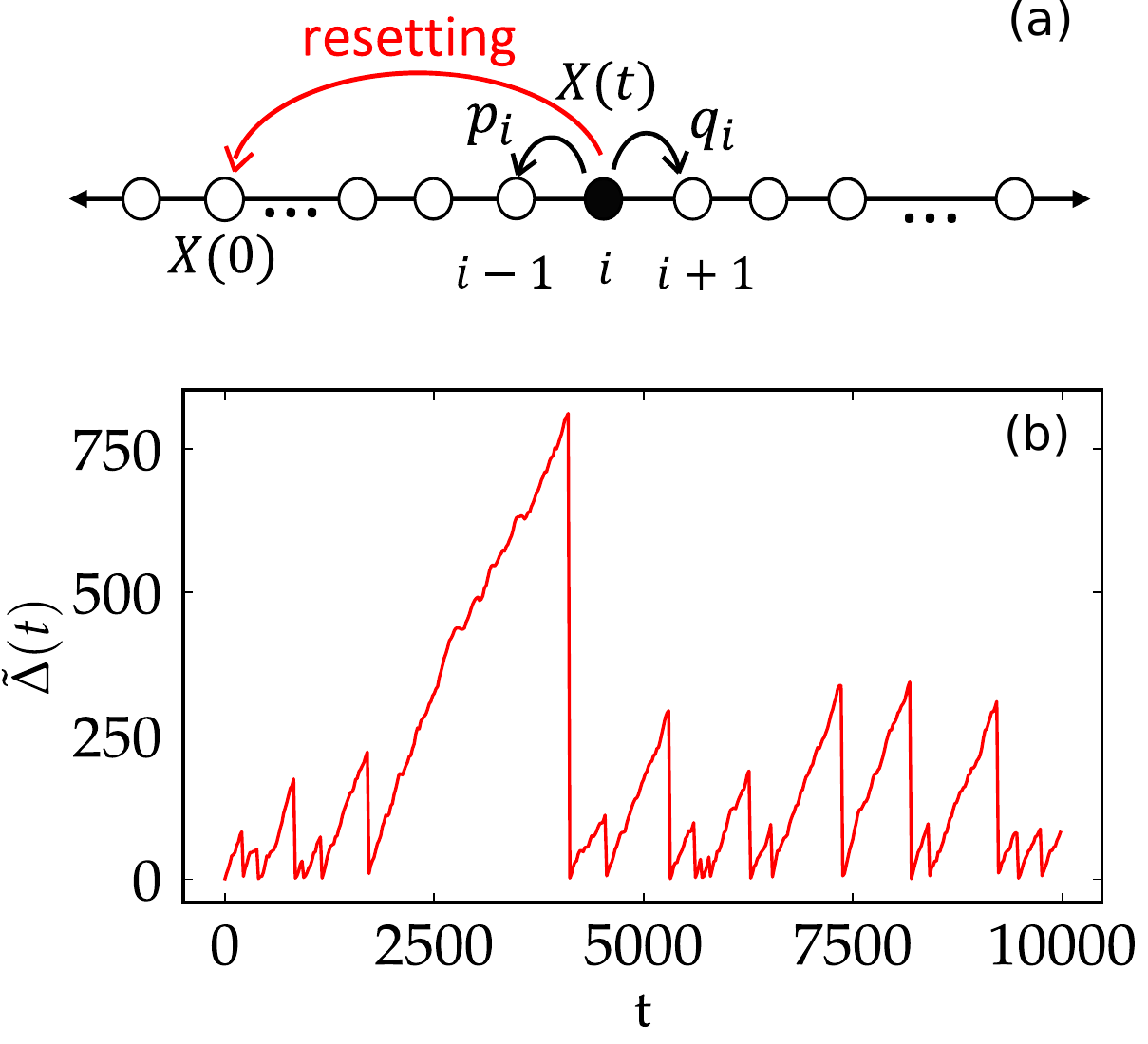}
\caption{(a) Schematic representation of our model: the particle on site $i$ at time $t$ hops at time $t+1$ to either the right or left-neighboring sites with site-dependent probabilities $q_i$ and $p_i=1-q_i$, respectively. In addition, stochastic resetting events (red arrow) reset the particle instantaneously to the origin or to a previously visited site.  (b) For a given realization of the disorder and an initial condition, the particle's trajectory exhibits alternating phases of growth and sudden resets to the origin. Here, $\tilde{\Delta}(t)$ represents the displacement of the particle with respect to the initial position.}
	\label{f1}
\end{figure}

We consider a one-dimensional model of a single particle hopping on an infinite lattice. The particle when on site $i$ at time $t$ can go at time $t+1$ either to site $i+1$ with probability $q_i$ or to site $i-1$ with the complementary probability $p_i = 1-q_i$ (Fig.~\ref{f1}(a)). 
The hopping probabilities $p_i$ (equivalently, $q_i$) are drawn independently for every site from a given distribution, and are quenched, that is, they remain fixed in values over time. In the absence of disorder, when all the $p_i$'s are the same, the model is well-known, leading in appropriate limits to a drift-diffusion evolution equation~\cite{fellerbook}. However, nontrivial behaviour emerges when the probabilities $p_i$ are quenched random variables, as in our case~\cite{alexander1981,derrida1983}. We draw the hopping probabilities from an algebraic distribution  
\begin{equation} \label{hoppingdist}
	\mathcal{P}(p_i)=\frac{\lambda}{p_l\left(\frac{p_i}{p_l}\right)^{1+\lambda}},
\end{equation}
where $\lambda$ is the scaling parameter, and $p_l>0$ is the smallest allowed value of $p_i$. The distribution has a rich behavior with varying $\lambda$: both the average and the variance diverge for $0<\lambda<1$, whereas both are finite for $2<\lambda< \infty$. In the intermediate regime, the average is finite, but the variance diverges. In general, $p_i$ has the range $[p_l,\infty]$. However, in our case, $p_i$ being a probability and not a rate, we take its values in the range $[p_l,1]$. 

To simulate an infinite system, we consider a system with $N$ sites, with $N$ large enough ($\sim 10^4$); we have checked that our results do not change on considering larger values of $N$. Once the hopping probabilities are assigned to the sites, we choose the initial position $X(0)$ of the particle to be one of $N$ sites with equal probability, and evolve the time-dependent particle position $X(t)$ up to $t=10^4$ via Monte-Carlo dynamics that follows the dynamical rules detailed above. For a given realization of the quenched random variables $\{p_i\}$, we define the displacement $\tilde{\Delta}(t)$ of the particle at time $t$ as $\tilde{\Delta}(t)\equiv|X(t)-X(0)|$, which we first average over $10^4$ distinct choices of the initial position, to obtain $\bar{\Delta}(t)=\langle \tilde{\Delta}(t)\rangle_0$. Finally, we average over $10^2$ independent realizations of the set $\{p_i\}$, to obtain the averaged displacement $\Delta(t)=\langle \bar{\Delta}(t)\rangle$.

To implement the resetting dynamics, we reset the particle position to its initial location (schematically shown by the red arrow in Fig. \ref{f1}(a)) over and over again after a time interval $\tau$ drawn from a given distribution. The particle displacement $\tilde{\Delta}(t)$ under resetting is shown in Fig. \ref{f1}(b). In this work, motivated by various physical problems, we have used three representative choices of the reset time distributions as we discuss below. Several analytical results are known for the particle displacement in the absence of resetting for this one-dimensional system; we present a brief overview of these results and their comparison with our simulations in Appendix~\ref{withoutresetting}. Below, we first present results for the dynamics inspired by microtubule growth.

\section{Results}\label{result}

\begin{figure}
	\includegraphics[width=8.6cm]{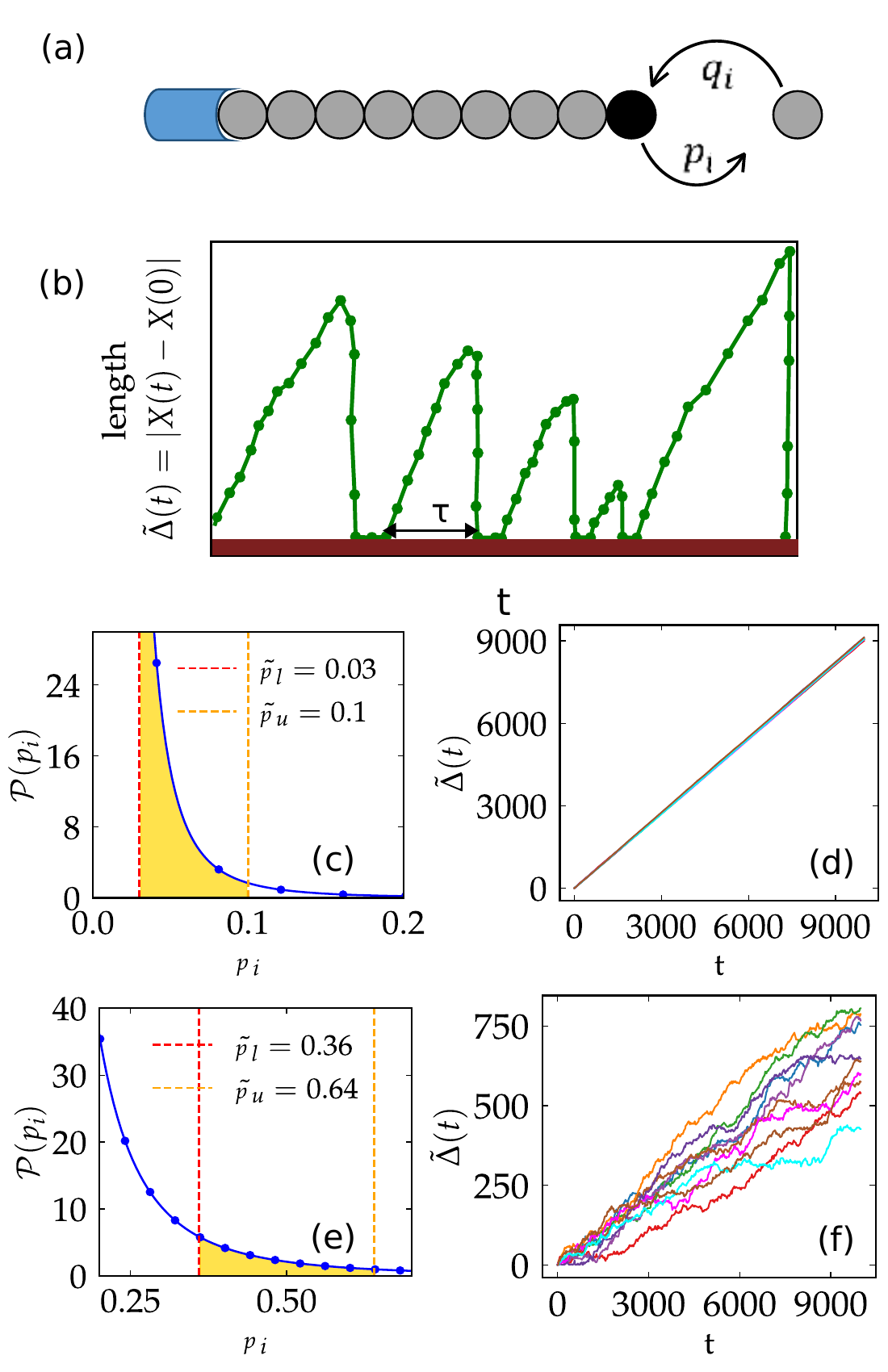}
	\caption{(a) Schematic of attachment-detachment kinetics of a tubulin unit in a mictotubule filament by probability $q_i$ and $p_i$, respectively. (b) Growth and catastrophe dynamics of microtubule length with time in the experiments of Ref.~\cite{gardner2011depolymerizing}. (c)The probability distribution function $P(p_i)$, defined over the interval $p_i \in [0.03, 0.1]$, shows a higher density of values near $p_i = 0.03$, indicating a strong bias in the lattice.
		(d) We plot the absolute displacement $\tilde{\Delta}(t)$ of a particle from its initial position with time for the biased case. Ten such trajectories are plotted together. (e) Probability distribution function, for the less biased case where the allowed values of $p_i \in [0.36,0.64]$. (f) Plot of absolute displacement $\tilde{\Delta}(t)$  as a function of time for ten representative trajectories in the presence of unbiased disorder.}
	\label{f2}
\end{figure}

\subsection{Resetting: microtubule-inspired dynamics}
\label{microtubule}

We apply the resetting formalism to a disordered system from the perspective of microtubule growth dynamics. Microtubules are dynamic long polymers spread across a cell, and are formed by polymerisation of GTP-bound Tubulin monomers. The dynamic end of the microtubule undergoes continuous addition and removal of monomers, leading to the overall growth of the microtubule \cite{mitchison1984dynamic}. These rates can depend both on length and other local properties of the monomers \cite{li2014theoretical,yadav2011length, howard2009growth, margolin2012mechanisms, hill1984introductory}. During the growth process, GTP-tubulin hydrolyzes to GDP-tubulin, a less stable form. A protective GTP cap stabilizes the microtubule. When the GTP hydrolysis rate is high, and the protective GTP cap is lost \cite{babinec1995self}, subunits disassemble rapidly, leading to microtubule shrinkage; this is known as catastrophe. In addition, polymerisation and depolymerisation processes depend on various factors, such as kinesin concentration, type and concentration of motor proteins~\cite{gardner2011depolymerizing}, exposure to shear flow, confinement~\cite{zelinski2012dynamics}, and environmental barriers~\cite{ambrose2011clasp}.

We model this growth process as particle hopping on a one-dimensional lattice and demonstrate that resetting dynamics can bring in crucial insights. As schematically shown in Fig. \ref{f2}(a), the particle position on the one-dimensional lattice represents the tip of the microtubule. Attachment of a subunit moves the tip one lattice unit to the right and detachment to the left. To account for the effects of biological sequence dependence and local environments on attachment-detachment rates, we treat them as quenched random variables. $q_i$ represents the probability of attachment and $p_i$, the detachment, as illustrated in Fig. \ref{f2}(a). We consider the unit of time in which one micro-event occurs. The length of the microtubule filament is given by $\tilde{\Delta}(t)$.

Figure \ref{f2}(b) shows the experimental data from Ref. \cite{gardner2011depolymerizing} for the growth dynamics of a microtubule as a function of time. As discussed above, the microtubule length grows with time and also undergoes intermittent `catastrophe' events \cite{howard2009growth,yadav2011length}. This `catastrophe' event is similar to resetting a particle to its initial position. The brown part in Fig. \ref{f2}(b) represents the substrate on which the sub-units assemble (similar to a nucleation centre), leading to the growth of the microtubule. There is a time gap between the catastrophe and the subsequent growth of the microtubule (Fig. \ref{f2}b). However, this waiting time is not crucial, as we measure the catastrophe time from the start of the growth process and are interested in the distribution of microtubule length. Therefore, we ignore this waiting time in the model (Fig. \ref{f1}a). 
Experiments indicate that microtubule catastrophe times follow a Gamma distribution \cite{odde1995kinetics,gardner2011depolymerizing}; accordingly, we draw our reset times from a gamma distribution
\begin{equation}\label{gamma_eqn}
	\mathcal{R}_n(t) = \frac{r^n t^{n-1} e^{-rt} }{\Gamma(n)}
\end{equation}
where $n$ is the shape parameter and $r$ is the rate parameter, and $\Gamma(n)$ is the Gamma function. We chose $n=3$ and $r=0.006$, so that the average reset time is $\langle t \rangle = {n}/{r} = 500$ time steps. We first study the dynamics with resetting the particle to the initial position, and then to a uniform position.

In the absence of a detailed knowledge of the disorder in the attachment-detachment probabilities of the sub-units, we consider two distinct cases. First, the attachment is strongly biased, and second, it is less biased. As discussed in Appendix \ref{withoutresetting}, we can control this by using a small and large $p_l$, respectively. However, to avoid the rare events of the stuck trajectories, discussed in the Appendix (Fig. \ref{f8}d), we also impose upper bounds on $p_i$. Accordingly, for the first case of the strongly biased case, we chose $p_i \in [0.03,0.1]$ with $\langle p_i\rangle = 0.046$. Here, the particle moves in a more steady, less fluctuating manner over time. In the other case, we choose $p_i \in [0.36,0.64]$ with $\langle p_i\rangle = 0.46 $ to ensure randomness with drift. We don't exceed the upper limit of $p_u=0.64$ to eliminate trapped trajectories. As expected, trajectories are strongly directed in the biased case [Figs. \ref{f2}(c) and \ref{f2}(d)]. By contrast, trajectories exhibit more stochastic fluctuations while retaining a net drift in the less biased case [Figs. \ref{f2}(e) and \ref{f2}(f)]. For both cases, we have used $\lambda=2.1$.

\subsection*{Reset length distribution}

{\it Resetting to initial position}:
Motivated by the experiments of Ref. \cite{gardner2011depolymerizing}, we first assume that the rapid shrinking of the microtubule after it loses the stabilizing cap disassembles the entire length. This corresponds to resetting the particle to its initial position. In the experiments, we know the distributions of the reset times and reset lengths, $l_r$ (microtubule lengths just before the resetting event). Although both distributions follow gamma distributions, Eq. (\ref{gamma_eqn}), the parameters are quite different. We first investigate how the nature of the $p_i$ affects the distribution of $l_r$.

\begin{figure}
	\includegraphics[width=8.6cm]{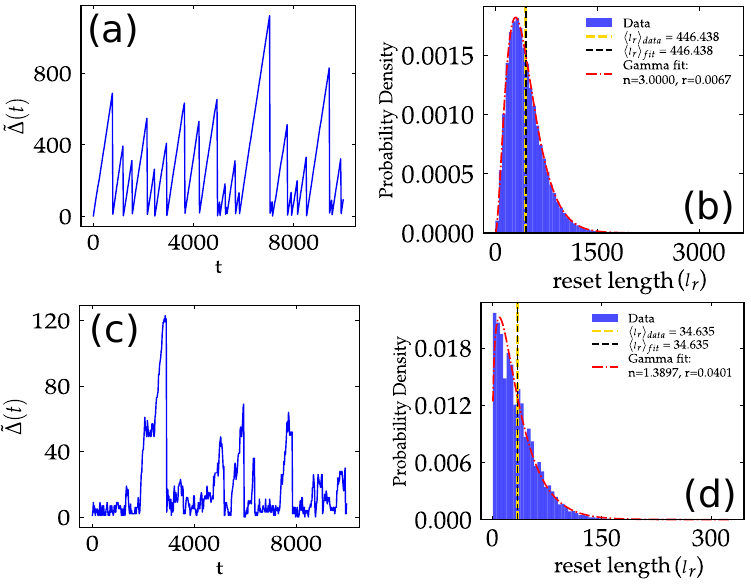}
	\caption{Displacement variation with and reset length distributions under stochastic resetting to the initial position. (a) Time evolution of particle displacement under stochastic resetting in the strongly-biased case ($p_i \in [0.03, 0.1]$). (b) Distribution of reset lengths for the strongly-biased case. The red dashed line indicates a Gamma fit with parameters $n = 3.00$ and $r = 0.0067$. (c) Displacement dynamics for the less-biased case ($p_i \in [0.36, 0.64]$) under stochastic resetting. (d) Reset length distribution for the less-biased case, fitted with a Gamma distribution (fit parameters: $n = 1.39$, $r = 0.04$).}
	\label{f3}
\end{figure}

We first consider the biased case, with $p_i\in [0.03,0.1]$. In this case, the trajectories are smooth, where the particle hops right most of the time and very rarely to the left (Fig. \ref{f3}a). 
We extracted $l_r$ as the absolute particle displacement from the initial position to a step before resetting for many trajectories with different initial positions and several realizations of the disorder. 
We find that $l_r$ follows a gamma distribution with step parameter $n = 3.00$ and $r=0.0067$, with the average reset length $\langle l_r \rangle \sim 446$ (Fig. \ref{f3}b). Note that $n$ and $r$ are very similar to those for the reset time distribution, Eq. (\ref{gamma_eqn}).
We can rationalize this as follows. When the bias is quite strong, the effects of disorder are not crucial for different trajectories. In that case, the velocity, $v$, remains constant and nearly 1; $l_r$ is given by the velocity multiplied by the reset time $(l_r=v\tau)$. Therefore, we expect the two distributions to be identical:
\begin{align}\label{lr_eqn}
	P(l_r) &= \int_0^\infty \mathcal{R}_n(\tau)\delta(l_r-v\tau)\d\tau=\int_0^\infty re^{-r\tau}\delta(l_r-v\tau)\d\tau \nonumber\\
	&=\frac{(r/v)^n l_r^{n-1} e^{-(r/v)l_r} }{\Gamma(n)}.
\end{align}

However, the distribution of $l_r$ in the experiments shows different values of $n$ and $r$ than those for the distribution of catastrophe time \cite{gardner2011depolymerizing}. We find that this behavior emerges when the bias is not very strong. We now present the results for the less biased case, where $p_i\in [0.36,0.64]$. The trajectories in this case are rugged, as shown in Fig. \ref{f3}(c). The distribution of $l_r$ differs from the gamma distribution. If we fit the data with a gamma distribution, the values of $n$ and $r$ become different from those for the reset time distribution: $n=1.39$ and $r=0.04$ (Fig. \ref{f3}d). The simple argument leading to Eq. (\ref{lr_eqn}) does not apply as $v$ depends on the local values of $p_i$ in this case. We also find $\langle l_r \rangle \sim 34$, that is, the catastrophe occurs at a much shorter average length of the microtubule. These results imply that the non-zero depolymerization of subunits plays a crucial role in microtubule growth dynamics.

{\it Resetting to a randomly chosen previously visited position:}
In many experimental scenarios, there are only a few GDP-bound tubulins at the tip, followed by a stabilizing GTP-bound tubulin inside. In such cases, only the GDP-bound segment depolymerizes, leaving the rest of the filament intact \cite{aumeier2016self,cassimeris1988real}.
In addition, another process, known as severing, occurs in growing microtubules and actin filaments. Specialized severing proteins control this process, in which they bind to cytoskeletal filaments and induce breakage, thereby fragmenting the filaments \cite{pavlov2007actin, roland2008stochastic, antal2007dynamics}. This event is rapid compared to the normal growth dynamics and can be viewed as a resetting process. During severing, the entire filament does not collapse; instead, only a finite part disassembles. Motivated by these examples, we next consider a resetting process in which a particle stochastically resets to a random position between its initial and current positions. Before each resetting event, we record the particle's displacement. We then draw a uniformly distributed random number within the interval defined by the initial and current positions, and reset the particle to that randomly chosen position.

\begin{figure}
	\includegraphics[width=8.6cm]{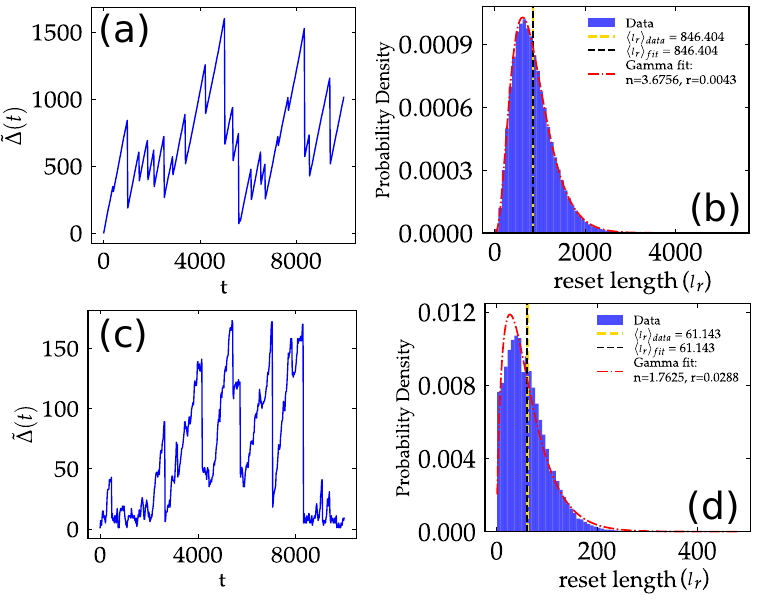}
	\caption{Displacement and reset length distributions under stochastic resetting to the uniform position. (a) Temporal evolution of particle displacement in the strongly biased regime ($p_i \in [0.03, 0.1]$) under stochastic resetting. (b) Corresponding reset length distribution in the strongly biased case, with a Gamma fit shown by the red dashed line ($n = 3.68$, $r = 0.0043$). (c) Displacement dynamics under stochastic resetting in the less biased regime ($p_i \in [0.36, 0.64]$). (d) Reset length distribution for the less biased case, fitted to a Gamma distribution with parameters $n = 1.76$ and $r = 0.029$.}
	\label{f4}
\end{figure}

We now investigate the effect of resetting to a randomly chosen previously visited position on the dynamics. For a quantitative comparison, we keep the other parameters constant and obtain the distribution of $l_r$ for the biased case, with $p_r\in [0.03,0.1]$, and for the less biased case, with $p_r\in [0.36,0.64]$. We show the trajectories for the two cases for a particular realization of the hopping probabilities in Figs. \ref{f4}(a) and (c) respectively. We again find that the distribution of $l_r$ remains a gamma distribution for the biased case, as in Fig. \ref{f3}(b); however, the values of $n$ and $r$ differ: $n=3.67$ and $r=0.0043$ with $\langle l_r \rangle \sim 846.4$ (Fig. \ref{f4}b). In addition, the distribution of $l_r$ for the less biased case also remains a gamma function (Fig. \ref{f4}d), although the parameters vary: $n=1.76$ and $r=0.029$. In this latter case, we find $\langle l_r \rangle \sim 61.1$. Thus, the parameters of the gamma function for the $l_r$ distribution differ from those of the reset time distribution for both the biased and the less-biased cases when the particle position is reset to a randomly chosen previously visited site.

\begin{figure}
	\includegraphics[width=8.4cm]{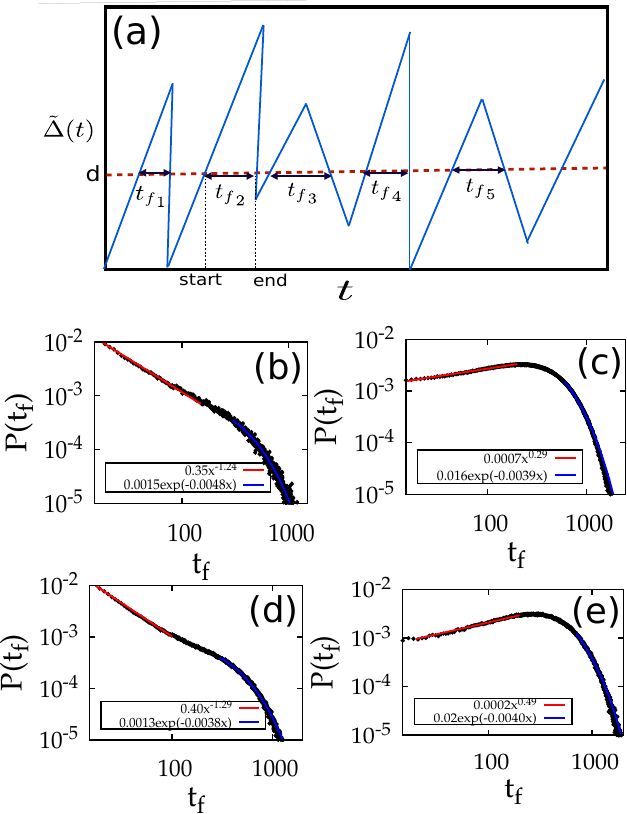}
	\caption{(a) Schematic representation of first passage times ($t_f$) for a particle at $d$. (b-e) Log-log plots of the first passage time distribution, $P(t_f)$, for $d=100$, shown for different resetting cases. In all the panels, we fit the initial part, i.e., the small time part, by a power-law $t_f^{a}$, and the long time part by $e^{-b t_f}$. (b) Less biased and resetting to the initial position: $a=-1.24$, $b=0.0048$ (c) Strongly biased and resetting to the initial position: $a=0.29$, $b=0.0039$(d) Less biased and resetting to a uniform position: $a=-1.29$, $b=0.0038$. (e) Less biased and resetting to a uniform position: $a=0.49$, $b=0.0040$.
    }
	\label{f5}
\end{figure}

\subsection*{First passage time distribution}

The first passage time (FPT) refers to the time a process takes to reach a specified state for the first time. It is a random variable and depends on the underlying dynamics of the process. A study by Needleman et. al \cite{needleman2010fast, mirny2010quantitative} tracked how long a single labeled monomer stays attached to the tip of a filament (single molecule lifetime in the filament). We can define this time interval from the time the monomer first joins the growing tip to when it leaves due to filament shrinkage. Since the filament’s length changes randomly, we treat this lifetime as FPT, the time it takes for the filament to return to the same length where the monomer was added, after random growth and shrinkage steps. The FPT distribution in the absence of disorder fits well with the function $f(t) \sim t^{-3/2} \exp(-t/\tau)$; we can obtain this functional form following the diffusion-drift model, where a particle diffuses on a one-dimensional lattice with a net bias towards a particular direction \cite{needleman2010fast}.

\begin{table}[htbp]
\begin{center}
  \caption{Comparison of first passage time distributions for initial and uniform position resetting}
\begin{tabular}{ | c | c| c| } 

  \hline
 \makecell{Resetting \\ position }& Strongly biased & Less biased \\ 
  \hline
  Initial & \makecell{Initially follows power \\ law $t_f^{a}$ with $a=0.29$ \\ and at large time \\ exponential behaviour \\ $e^{-b t_f}$ with  $b=0.0039$. \\This behavior is \\ similar to the resetting \\ time distribution.} & \makecell{Decreases at short \\ times as a power law\\ $t_f^{-a}$ with $a=1.24$.  \\ Exponential cut-off \\ at large times\\ $e^{-bt_f}$ with $b=0.0048$.} \\ 
  \hline
  Uniform & \makecell{At short times \\ increases as a power law \\ $t_f^a$ with $a=0.49$ \\ and exponential cut-off\\ at large times \\ $e^{-bt_f}$ with  $b=0.0040$. }  & \makecell{Decreases at short times \\ as a power law\\ $t_f^{-a}$ with $a=1.29$. \\ Long time behavior\\ is exponential \\ $e^{-bt_f}$ with $b=0.0038$.} \\ 
  \hline
\end{tabular}
\end{center}
\end{table}

Following Ref. \cite{needleman2010fast}, we also define the first passage time as follows: We consider a particular displacement $d$ designated by a specific site on the lattice, as shown for a schematic trajectory in Fig. \ref{f5}(a). We start the timer when the particle reaches $d$ and record the first-passage time when it returns to $d$ for the first time. This time also signifies how long the particle spends beyond the displacement $d$. Figure \ref{f5}(a) shows five such FPT, where $t_{f_1}, t_{f_2},t_{f_4}$ are contributions from resetting and $t_{f_3}, t_{f_5}$ are from hopping. In our system, the FPT will have contributions from both hopping dynamics and resetting.

For the less biased case, although the general forms of the distribution remain the same, a power law followed by exponential cut-off, the exponents depend on the nature of disorder and the resetting protocol. For the less-biased cases, Figs. \ref{f5}(b) and (d) show the distributions, $P(t_f)$, for the reset to the initial position and to a randomly chosen previously visited position, respectively. The power-law decay at early time remains similar in both cases, with the exponent $\sim 5/4$. However, the exponential cut-off is slower in the latter case because a reset does not always result in FPT when the chosen reset position exceeds $d$.

By contrast, in the biased case, $P(t_f)$ exhibits different behavior: the particle moves nearly linearly before the reset, and $t_f$ is nearly equal to the reset time. Therefore, the distribution of $t_f$ is similar to the Gamma distribution. Figures \ref{f5}(c) and (e) show the distributions for the reset to the initial position and to a random position, respectively. We have checked that varying $d$ does not qualitatively change the distributions.

\begin{figure}
	\includegraphics[width=8.6cm]{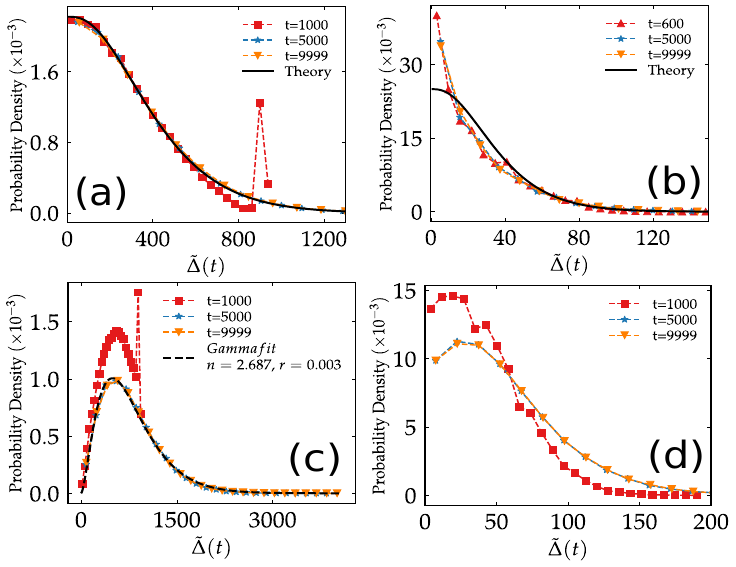}
	\caption{The probability distribution of displacement, $\tilde{\Delta}(t)$, at different times.  (a) Strongly biased and resetting to initial position, (b) less biased and resetting to initial position, (c) strongly biased and resetting to random position, (d) less biased and resetting to random position. In all cases, the distribution converges to the steady state after a finite time. Solid lines in (a) and (b) represent the analytical expression, Eq. \ref{displ_eqn}. }
	\label{f6}
\end{figure}

\subsection*{Steady state distributions due to resetting}

We now investigate the steady state imposed by resetting, specifically, we study the steady-state probability distribution of displacement $(\tilde{\Delta}(t))$. Though there is a net bias in the particle's direction of motion due to our selection of $p_i$, resetting leads to a steady state (independent of $t$). Such emergence of the steady-state due to resetting has been widely studied \cite{evans2011,gupta2022stochastic}.

Figures \ref{f6}(a) and (b) show the distributions of $(\tilde{\Delta}(t))$ at various times for the resetting to initial position with the strongly-biased ($p_i\in [0.03,0.1]$) and the less-biased ($p_i\in [0.36,0.64]$) cases, respectively. 
Similarly, Figs. \ref{f6}(c) and (d) show the distributions at various times for the strongly-biased and the less-biased cases when we reset the particle to randomly chosen positions. The peaks in the early-time distributions for the strongly-biased cases are due to the following: the velocity is nearly 1, as most particles move to the right most of the time. The resetting is not very effective at the early times, particularly for $t=1000$. For many trajectories, the particle is found around $\tilde{\Delta}(t)=1000$, giving rise to these peaks in the distributions. Over time, resetting brings the particle back to its initial position (or to a random position for random resets), and the peak disappears. As the velocity is much smaller for the less biased case, when the particle has a significant probability to hop left, this peak is not visible (Figs. \ref{f6} b and d).

Assuming the bias is strong and the particle's position is reset to its initial value, we can obtain the analytical form of the steady-state distribution. For this, we must know the survival probability, which is the probability that the particle's last reset occurred exactly $t$ times ago. If the reset time distribution is $\mathcal{R}_n(\tau)$, we obtain the normalized survival probability as $S(t)=\left(1-\int_{0}^{t}\mathcal{R}_n(t')dt'\right)/\langle \tau\rangle$. If the velocity of the particle is $v$, then the distance traversed in $t$ is $\tilde{\Delta}= vt$. Therefore, similar to the distribution of $l_r$ [Eq. (\ref{lr_eqn})], using Eq. (\ref{lr_eqn}), we have
\begin{align}\label{displ_eqn}
	P(\tilde{\Delta}) &= \int_0^\infty S(t)\delta(\tilde{\Delta}-v t)dt= \frac{r}{v n}\,\frac{\Gamma\left(n, \frac{r}{v}\tilde{\Delta}\right)}{\Gamma(n)},
\end{align}
where $\Gamma(n,\frac{r}{v}\tilde{\Delta})$ is the incomplete Gamma function. As discussed earlier, this result applies only to the strongly biased case. Indeed, considering $v \sim \langle p_i \rangle - \langle q_i \rangle \sim 0.9$, we obtain good agreement of Eq. (\ref{displ_eqn}) with the simulation data (Fig. \ref{f6}a). By contrast, $v$ depends on the local environment for the less-biased case, and considering the average $v\sim 0.08$, Eq. (\ref{displ_eqn}) does not agree well with the simulation data (Fig. \ref{f6}b).

Figure \ref{f6} also shows that when resetting to the initial position, the probability distribution in the less-biased case reaches a steady state earlier than that in the biased case. However, for resetting to a random position, both the biased and less-biased systems reach steady state at similar times. The steady-state distributions do not follow the gamma distribution except for the strongly biased case with resetting to a random position (Fig. \ref{f6} c); the dashed line shows the fit with a gamma function with fitting parameters $n=2.68$, $r=0.0034$ (for $t=9999$).

\subsection{Other resetting time distributions}
\label{otherreset}
We next discuss the effects of other resetting-time distributions on the particle's dynamics in the presence of quenched disorder. We first consider an exponential distribution and then a power-law distribution.\\

\begin{figure}
\includegraphics[width=8.6cm]{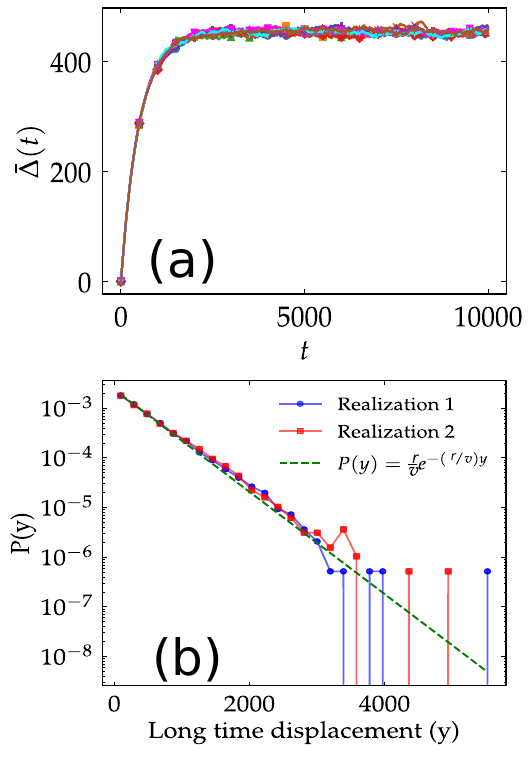}
\caption{For biased disorder case: (a) mean displacement of particle when exponential resetting($r=0.002$) for 10 different realizations. (b) Simulation data of long-time displacement probability (red and blue plots) match well with the theoretical function, Eq. (\ref{pofy}), shown by the green dashed line. }
\label{exponential_resetting}
\end{figure}

{\bf Exponential resetting}:
Exponential resetting in a diffusive system can induce a non-equilibrium steady state \cite{evans2011, gupta2022stochastic}. Several studies investigated the impact of resetting in heterogeneous media, revealing its ability to suppress trapping \cite{barbini2024lattice}, tune relaxation behavior on non-trivial dielectric response \cite{petreska2022tuning}, induce localization transitions \cite{boyer2019anderson}, or modulate the first-passage behavior \cite{lenzi2022transient}, depending on the nature of disorder. Our system differs from these studies in that it exhibits quenched disorder, in which the steady state depends on the particular realization of the disorder. 

We consider the same system as before, with the strongly-biased ($p_i\in[0.03.0.1]$) and less-biased ($p_i\in[0.36,0.64]$) distributions of the hopping probabilities. We reset the particle's position to its initial position after a waiting time drawn from an exponential distribution with rate $r$. Thus, the probability of a reset occurring at time $t$ is
\begin{equation*}
	\mathcal{R}(t)=re^{-rt}
\end{equation*} 
where $\langle t \rangle=1/r$. For our simulations, we use $r=0.002$, so that the average resetting time $\langle t\rangle=500$ remains the same as before.

We show the initial position averaged displacement $\bar{\Delta}(t)$ for specific realizations of the disorder for the biased case in Fig. \ref{exponential_resetting}(a). The long-time displacement indeed reaches a steady state that depends on the specific realization of the disorder. The steady-state appears on a timescale of the order of the resetting time. We calculate the distribution of the long-time displacement $y=\tilde{\Delta}({t\to\infty})$ and plot $P(y)$ as a function of $y$ in Fig. \ref{exponential_resetting}(b) for two different disorder realizations. Following the same argument as in Eq. (\ref{displ_eqn}), we can calculate the distribution of $y$ as
\begin{align}\label{pofy}
	P(y) = \f{r}{v}e^{-ry/v}.
\end{align}
The mean of the average long-time displacement $\langle y\rangle=v/r$. Figure \ref{exponential_resetting}(b)  shows the simulation results by symbols, and the line is the analytical expression, Eq. (\ref{pofy}). The agreement is quite good. Note that, for the exponential resetting, the normalized survival probability, $S(t)$, and the reset time distribution, $\mathcal{R}(t)$, are the same. Therefore, $P(l_r)$ and $P(y)$ follow the identical distributions.

For the less biased case, $P(y)$ and $P(l_r)$ do not remain exponential. Moreover, here the particle moves less before being reset because the velocity is less. The average reset length $\langle l_r \rangle \sim 434.0$ in biased disorder and $\langle l_r \rangle \sim 33.4$ in the unbiased case.\\

\begin{figure}
	\includegraphics[width=7 cm]{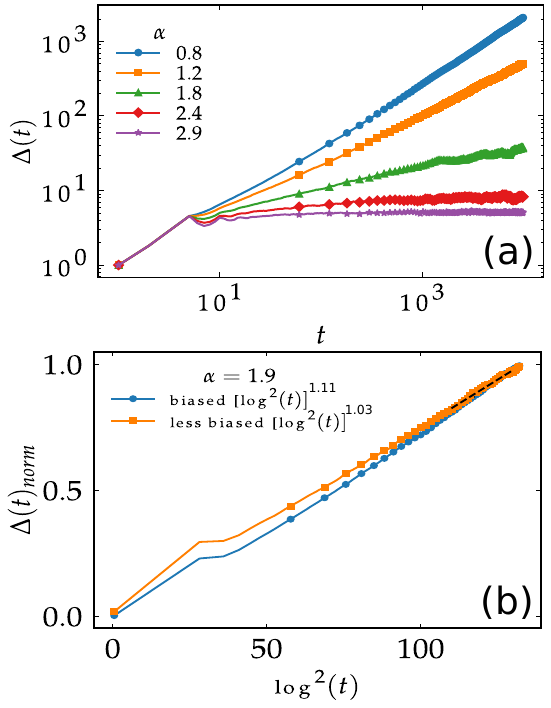}
	\caption{(a) Mean displacement of particle for different values of $\alpha$ for the strongly biased case. (b) The normalized mean displacement $\Delta(t)_\text{norm}$ follows linear behavior with $\log^2(t)$ at long times for $\alpha=1.9$ for both biased and less biased disorder cases.}
	\label{powerlaw}
\end{figure}

{\bf Power law resetting}:
How will a broader resetting distribution affect the particle dynamics? To answer this, we chose the reset time distribution as a power law,
\begin{equation*}
	\mathcal{R}(t)=\f{\alpha}{T_0\left(\f{t}{T_0}\right)^{1+\alpha}}
\end{equation*}
where $\alpha > 0$ is the resetting exponent that controls the heaviness of the tail. A smaller $\alpha$ gives a heavier (slower-decaying) tail, meaning long waiting times between resets are more probable. $T_0$ is the lower cutoff of the distribution. We chose $T_0 = 200$ for our simulation so that $\langle t \rangle$ remains comparable to previous cases, for example, $\langle t \rangle\sim 500$ for $\alpha=1.8$. 
When the distribution is broad (for smaller $\alpha$), the probability of large values of the reset time is significant. In that case, the dynamics do not have a steady state due to the rare but dominant long trajectories without any resetting \cite{eule2016non, gupta2019stochastic}.

We study the evolution of mean displacement ($\Delta(t)$) with time for different $\alpha$ values for the biased ($p_i\in[0.03,0.1]$) and the less biased ($p_i\in[0.36,0.64]$) cases. For both cases of disorder, we find a steady state when $\alpha$ is large, as shown in Fig. \ref{powerlaw}(a) for the biased case. In general, the biased case shows larger values of $\Delta(t)$ for the same disorder and $\alpha$. 
$\Delta(t)$ continues to grow at lower $\alpha$ for both cases. 

For this specific type of resetting, we can obtain an interesting type of growth law, $log^2t$, that is sometimes observed in systems with disorder and is known as the Sinai law. Examples include the dynamic heterogeneity (DH) length scale in disordered glassy systems with pinned particles \cite{saroj2014}, DNA unzipping \cite{lubensky2002}, etc.
Within our model, the specific value of $\alpha$ showing this Sinai law depends on $T_0$; We find that $\Delta(t)$ grows as $\log^2t$ for $\alpha=1.9$ and $T_0=200$ for both cases of disorder. We show the results for both the biased and less biased cases in Fig. \ref{powerlaw}(b); we plot the renormalized values of the displacement, $\Delta_\text{norm}(t)$, to show them in the same figure.

\begin{figure}
	\includegraphics[width=8.6cm]{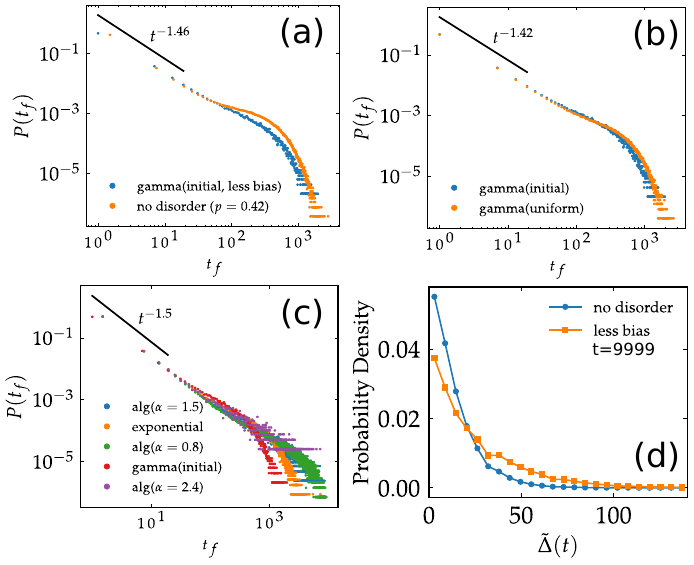}
	\caption{Effects of disorder and various types of resetting protocols. (a) In the presence and absence of disorder in the less-biased case, for the gamma resetting time distribution and reset to the initial position. (b) For the less-biased disorder system with gamma-distributed reset times, reset to the initial state and to a randomly chosen position. (c) A less-biased disordered system, but with various resetting time distributions and resetting to the initial position. (d) Steady state probability densities for displacements in the presence and absence of disorder under the gamma-distributed reset time.
	}
	\label{comparison}
\end{figure}

\begin{table*}
	\centering
	\caption{Comparison of different resetting cases in the absence and presence of disorder}
	\label{table}
	\begin{tabular}{|c|c|c|c|c|}
		\hline
		\multirow{2}{*}{Properties}
		& \multirow{2}{*}{\makecell{Resetting time distribution \\ \textit{(initial position reset)}}}
		& \multirow{2}{*}{No disorder}
		& \multicolumn{2}{c|}{Disorder} \\
		\cline{4-5}
		& & & Less biased & Strongly biased \\
		\hline
		\multirow{3}{*}{\makecell{Steady state \\ distribution}}
		& Gamma
		& Not gamma dist. & Not gamma dist. & Not gamma dist. \\
		\cline{2-5}
		& Exponential
		& Not exp.(for $p=q$: Exponential ~\cite{evans2011} ) & Not exponential & Exponential \\
		\cline{2-5}
		& Algebraic($\alpha=0.8$)
		& No steady state & No steady state & No steady state \\
		\hline
		\multirow{3}{*}{\makecell{Reset length \\ distribution}}
		& Gamma
		& Not gamma dist. & Not gamma dist. & \makecell{Gamma distribution, \\ Same $n$ and $r \sim r/v$, \\ $v$ is velocity} \\
		\cline{2-5}
		& Exponential
		& Not exponential & Not exponential & Exponential \\
		\hline
	\end{tabular}
\end{table*}

\subsection{Role of disorder and comparison with different types of resetting}
\label{roleofdisorder}

We now elucidate the role of disorder in the system properties and how different types of resetting affect them. First, we show the differences in the distribution, $P(t_f)$, of the first passage time, $t_f$, in the presence and absence of disorder (Fig. \ref{comparison}a). We show the comparison for the less-biased case, with the reset time drawn from a gamma distribution and the particle reset to its initial position. On the other hand, in the absence of disorder, we set all hopping rates to the same value, $p_i=p=0.42$, which is the average for the system with disorder. Figure \ref{comparison}(a) shows that the initial part of the distribution remains the same for both systems, although the exponential decay part at larger $t_f$ differs. A very similar trend also appears when we reset the particle to the initial position or a uniform position (Fig. \ref{comparison}b). In addition, we have compared the distribution $P(t_f)$ for different reset-time distributions. Figure \ref{comparison}(c) shows $P(t_f)$ for various such distributions when we reset the particle to its initial position, we again find that the initial power law decay remains the same, $t_f^{-3/2}$, whereas the time-scale $\tau$ of the late exponential decay, $\exp[-t_f/\tau]$, differs. Similarly, the qualitative characteristics with the various resetting time distributions are also similar in the strongly biased case.

We have also found that disorder affects the steady state. For example, Fig. \ref{comparison}(d) shows the differences of the probability densities of $\tilde{\Delta}(t)$ in the presence and absence of disorder (as above) when the reset time has a gamma distribution and we reset the particle to its initial position. 

We have listed the differences for the three types of reset time distributions, and for the presence or absence of disorder, in the steady-state distribution (if it exists) and the reset length distribution, in Table \ref{table}.

\section{Discussion and Conclusion}
\label{disc}

To conclude, we have applied the resetting framework for systems with quenched disorder. We considered a well-known model of particle hopping with quenched probabilities to move right or left, where we obtained site-dependent hopping rates from an algebraic distribution and studied the dynamics under various resetting protocols. Depending on the values of the hopping probabilities, we considered two distinct cases: a strongly biased case, where the particle moves in a specific direction most of the time, and a less biased case, where the particle has a significant probability of moving in the other direction as well. As a concrete example, we have applied the resetting framework to the growth dynamics of microtubules. Motivated by the experimental data of Ref. \cite{gardner2011depolymerizing}, we first considered gamma-distributed resetting times and studied the distribution of reset lengths. Our study shows that the experimentally observed resetting-length distribution arises in the less-biased case. This result implies that disassembly events play a significant role in microtubule growth dynamics.

There are many systems in which resetting dynamics plays a crucial role; for example, the plastic yielding dynamics in glassy systems \cite{ozawa2018}, avalanches during earthquakes \cite{brace1966,vere1988variance}, etc. The resetting framework can provide crucial insights into these systems. But these systems are characterized by quenched disorder, and therefore, it is essential to extend the resetting phenomenology to systems with quenched disorder. Our work is a step in that direction. Of course, a lot needs to be analyzed to make these physical processes amenable to theoretical treatment under the resetting phenomenology, for example, the nature of the disorder and the resetting times. We have shown, within our simplified framework, that quenched disorder does exhibit its signatures even in the presence of resetting; a deeper understanding of these effects will allow us to understand the dynamics in more realistic systems. One crucial aspect that future studies must also consider is the spatial dependence of resetting events, as a plastic event makes it easier to induce another in the surrounding region, leading to avalanches \cite{antonaglia2014bulk}.

Resetting can lead to anomalously slow dynamics \cite{evans2020}. For example, we have shown that when the resetting time has a power law distribution, the displacement can grow as $\log^2t$; a similar growth law is observed for Sinai diffusion \cite{sinai1982lecture}. 
There are several directions in which one can extend the current study. For example, the differences between quenched vs annealed disorder and the use of different types of distributions to draw the site-dependent hopping probabilities. Another interesting direction will be to consider the conserved mass aggregation (CMA) and related mass-transport models, for instance by allowing fragmentation at disordered sites \cite{jain2001phases}, site-dependent hopping rates \cite{jain2003dynamics,juhasz2005partially}, or random site capacities \cite{gupta2015condensate}. On the other hand, stochastic resetting has recently been applied to interacting particle systems such as the zero-range process \cite{grange2024local} and spin models \cite{magoni2020ising}. It will be interesting to combine these two ingredients-quenched disorder and resetting-in the CMA framework to investigate possible new critical points, altered condensation transitions, and novel scaling behaviors.

\appendix
\setcounter{equation}{0}
\setcounter{figure}{0}

\renewcommand{\theequation}{A\arabic{equation}}
\renewcommand{\thefigure}{A\arabic{figure}}

\begin{figure}
	\includegraphics[width=8.6cm]{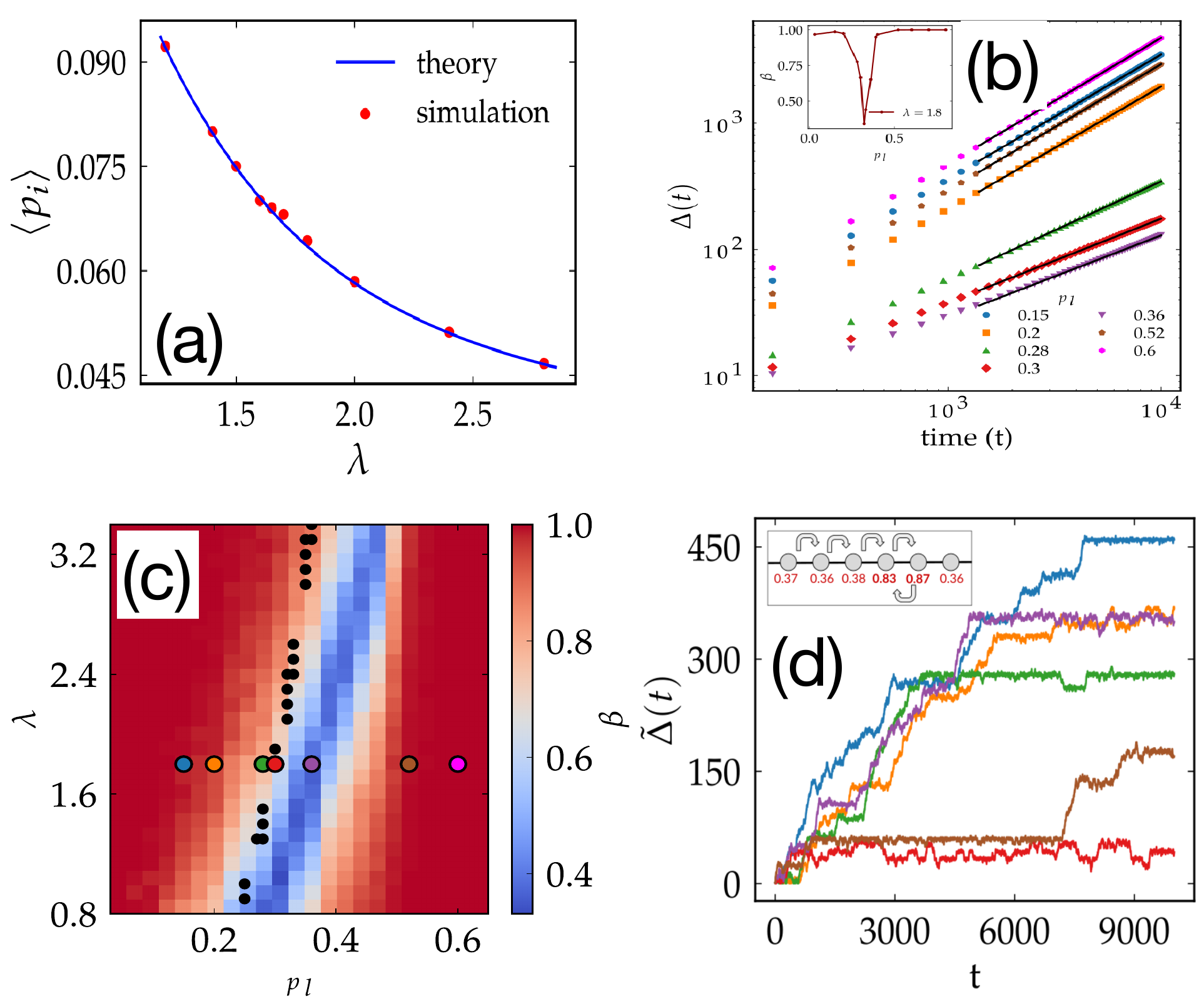}
	\caption{Comparison of our simulation with analytical results. (a)$\langle p_i \rangle$ for different $\lambda$ and $p_l=0.03$ in theory and simulation. (b) Log-log plot of $\Delta(t)$ vs time for different $p_l$. At long times, $\Delta(t)$ shows a power law relation ($\Delta(t) \sim t^{\beta}$). The inset shows the change of exponent ($\beta$) with the lower cut-off $p_l$. (c) The color plot shows how $\beta$ varies with $p_l$ and $\lambda$. The colored points represent the state points corresponding to the exponents in the inset of figure (b), and the black points correspond to $\langle [(1-p_i)/p_i]^\beta\rangle \sim 1$. This shows crossover from Eq. (\ref{case1_hopmodel}) to Eq. (\ref{case2_hopmodel}) and back. (d) Individual trajectories of particles in the power law regime ($p_i  \in [0.36,0.9]$). We observed stuck trajectories due to high hopping rates at a few sites(inset).}
	\label{f8}
\end{figure}

\section{Dynamics of a particle on a disordered lattice}
\label{withoutresetting}

Several analytical results are known for the dynamics of the particle in the presence of quenched disorder, but without resetting~\cite{sinai1982lecture,derrida1982, bernasconi1982}. In the specific case when the condition $\langle \ln p_i\rangle=\langle\ln q_i\rangle$ is satisfied, with $\langle ... \rangle$ representing the average over the distribution of $p_i$, one obtains $\Delta(t) \sim(\ln t)^2$~\cite{sinai1982lecture}. For the general case of hopping rates that do not satisfy the above condition, one finds~\cite{sinai1982lecture,solomon1975,derrida1983,bernasconi1982}\begin{align}\label{case1_hopmodel}
	\Delta(t)\simeq 
	\begin{cases}
		\left\langle \f{1}{p_i}\right\rangle^{-1}\left(1- \left\langle \f{q_i}{p_i}\right\rangle\right)t, \,\,\, \text{if} \left\langle \f{q_i}{p_i}\right\rangle<1, \\
		-\left\langle \f{1}{q_i}\right\rangle^{-1}\left(1- \left\langle \f{p_i}{q_i}\right\rangle\right)t, \,\,\, \text{if} \left\langle \f{p_i}{q_i}\right\rangle<1.
	\end{cases}
\end{align}
On the other hand, in the specific case when $\langle [(1-p_i)/p_i]^\beta\rangle=1$, with $\beta$ being a real number, one finds a power law increase of $\Delta(t)$:
\begin{align}\label{case2_hopmodel}
	\Delta(t)\sim \begin{cases}
		t^\beta \,\,\, &\text{if}\,\, \beta>0\\
		-t^{-\beta}\,\,\, &\text{if}\,\, \beta<0.
	\end{cases}
\end{align}

The above results are valid for a generic system with quenched disorder. We now show the comparison of these results with our simulations where $p_i$ follows the power-law distribution $\mathcal{P}(p_i)$ (Eq. \eqref{hoppingdist}). For $p_i \in [p_l,1]$, the mean is
\begin{equation}\label{mean}
	\langle p_i \rangle = \frac{\lambda}{\lambda-1}[p_l- {p_l}^\lambda].
\end{equation}
For $p_l=0.03$, Fig.~\ref{f8}(a) shows the comparison of $\langle p_i\rangle$ of the simulation results (symbols) and the analytical form (line) for various $\lambda$.
We find that $\langle p_i/q_i\rangle$ remains less than one for the particular value of $p_l=0.03$, and as predicted by Eq. (\ref{case1_hopmodel}),

$\Delta(t)$ grows linearly with $t$. However, as we increase $p_l$, $\Delta(t)$ starts showing power law behavior (Eq. \ref{case2_hopmodel}). Figure \ref{f8}(b) shows such behaviors of $\Delta(t)$ for several values of $p_l$ and $\lambda=1.8$.
To quantify this behavior in detail, we calculate the exponent $\beta$ by fitting $\Delta(t) \sim t^\beta$. In addition, we evaluate $\langle {q_i}/{p_i}\rangle$, $\langle {p_i}/{q_i}\rangle$, and $\langle [(1-p_i)/p_i]^\beta\rangle$. 
We find that, when $p_l \sim 0.03$,  $\langle {p_i}/{q_i}\rangle < 1$ and $\beta \sim 1$, consistent with Eq. \ref{case1_hopmodel}. When we increase $p_l$ (say around 0.33), we find that both $\langle {p_i}/{q_i}\rangle$ and $\langle {q_i}/{p_i}\rangle$ greater than 1, concomitantly $\langle [(1-p_i)/p_i]^\beta\rangle \sim 1$. In this case, we find $\beta$ to be much smaller than 1, in agreement with Eq. \ref{case2_hopmodel}. On further increase of $p_l$, $\langle {q_i}/{p_i}\rangle$ becomes less than one and $\beta$ rises again to 1, as in Eq. (\ref{case1_hopmodel}). This behavior with increasing $p_l$ persists for all values of $\lambda$ that we have explored in this study. 
Figure \ref{f8}(c) shows a color plot of $\beta$ on the two-dimensional $p_l$-$\lambda$ plane. The black dots in the plot correspond to $\langle [(1-p_i)/p_i]^\beta\rangle \sim 1$, which align with the minimum values of $\beta$ and highlight the crossover region.

Since $\mathcal{P}(p_i)$ is a power law distribution, most of the values of $p_i$ is concentrated around $p_l$. Thus, when $p_l$ is small, $p_i$ is mostly small and the particle is more probable to hop right, leading to the linear growth of $\Delta(t)$. On the other hand, when $p_l$ is larger, there are significant probabilities for the particle to move left. This is the regime where we see $\beta$ significantly less than unity (Fig. \ref{f8}c). To understand the origin of the power-law behavior of $\Delta(t)$, we examine the individual trajectories, (Fig. \ref{f8}d) where we chose $p_l \sim 0.3$. Interestingly, we observe that some trajectories show particle trapping. When $p_l$ is large, there is a significant probability for $p_i$ to be very close to unity. Till this site, the particle keeps moving towards it as $p_i$ for these earlier sites are still small. However, as it reaches this specific site, it is more probable to move left. Thus, such a site works like a reflecting wall and the particle remains stuck for a large time, of the order of $1/(p_i-1)$. Such a site rarely appears for smaller $p_l$. For even larger $p_l$, the particle is more probable to hop right and the power law behavior again disappears.

\section*{Acknowledgments}
We acknowledge the support of the Department of Atomic Energy, Government of India, under Project identification No. RTI 4007. The work of SG was supported by the Department of Atomic Energy, Government of India, under Project Identification Number RTI-4012; he thanks Ajay Salve and Kapil Ghadiali for their computational support.

%

\end{document}